\documentclass[aps,prb,twocolumn,groupedaddress,amssymb,superscriptaddress,showpacs,10pt]{revtex4}
\usepackage[pdftex]{graphicx} 
\usepackage{graphicx}
\usepackage{color}
\definecolor{dg}{rgb}{0,0.5,0} 
\usepackage{nicefrac}
\newcommand{\cefaof}{Ce\-Fe\-As\-O$_{1-x}$\-F$_{x}$}

\begin{document}
\title{Long- to short-range magnetic order in fluorine-doped Ce\-Fe\-As\-O }
\author{T.~Shiroka}
\affiliation{Laboratorium f\"ur Festk\"orperphysik, ETH-H\"onggerberg, CH-8093 Z\"urich, Switzerland}
\affiliation{Paul Scherrer Institut, CH-5232 Villigen PSI, Switzerland}
\author{G.~Lamura} 
\email[Corresponding author: ]{gianrico.lamura@spin.cnr.it}
\affiliation{CNR-SPIN and Universit\`a di Genova, via Dodecaneso 33, I-16146 Genova, Italy}
\author{S.~Sanna}
\affiliation{Dipartimento di Fisica ``A.\ Volta'' and Unit\`a CNISM di Pavia, I-27100 Pavia, Italy}
\author{G.~Prando}
\affiliation{Dipartimento di Fisica ``A.\ Volta'' and Unit\`a CNISM di Pavia, I-27100 Pavia, Italy}
\affiliation{Dipartimento di Fisica ``E. Amaldi'', Universit\`a di Roma 3 and CNISM, I-00146 Roma, Italy}
\author{R.~De Renzi}
\affiliation{Dipartimento di Fisica and Unit\`a CNISM di Parma, I-43124 Parma, Italy}
\author{M.~Tropeano}
\affiliation{CNR-SPIN and  Dipartimento di Fisica dell'Universit\`a di Genova, via Dodecaneso 33, I-16146 Genova, Italy}
\author{M.~R.~Cimberle}
\affiliation{CNR-IMEM and  Dipartimento di Fisica dell'Universit\`a di Genova, via Dodecaneso 33, I-16146 Genova, Italy}
\author{A.~Martinelli}
\affiliation{CNR-SPIN and  Dipartimento di Chimica dell'Universit\`a di Genova, via Dodecaneso 31, I-16146 Genova, Italy}
\author{C.~Bernini} 
\affiliation{CNR-SPIN and Universit\`a di Genova, via Dodecaneso 33, I-16146 Genova, Italy}
\author{A.~Palenzona}
\affiliation{CNR-SPIN and  Dipartimento di Chimica dell'Universit\`a di Genova, via Dodecaneso 31, I-16146 Genova, Italy}
\author{R.~Fittipaldi}
\affiliation{CNR-SPIN and  Dipartimento di Fisica dell'Universit\`a di Salerno, I-84084 Salerno, Italy}
\author{A.~Vecchione}
\affiliation{CNR-SPIN and  Dipartimento di Fisica dell'Universit\`a di Salerno, I-84084 Salerno, Italy}
\author{P.~Carretta}
\affiliation{Dipartimento di Fisica ``A.\ Volta'' and Unit\`a CNISM di Pavia, I-27100 Pavia, Italy}
\author{A. S. Siri}
\affiliation{CNR-SPIN and  Dipartimento di Fisica dell'Universit\`a di Genova, via Dodecaneso 33, I-16146 Genova, Italy}
\author{C.~Ferdeghini}
\affiliation{CNR-SPIN and  Dipartimento di Fisica dell'Universit\`a di Genova, via Dodecaneso 33, I-16146 Genova, Italy}
\author{M.~Putti}
\affiliation{CNR-SPIN and Dipartimento di Fisica dell'Universit\`a di Genova, via Dodecaneso 33, I-16146 Genova, Italy}
\date{\today}
\begin{abstract}
The evolution of the antiferromagnetic order parameter in \cefaof\ as a function of the fluorine content $x$ 
was investigated 
primarily via zero-field muon-spin spectroscopy. The long-range magnetic order observed in the undoped compound gradually turns into a short-range order at $x=0.04$, seemingly accompanied/induced by a drastic reduction of the magnetic moment of the iron ions. Superconductivity appears upon a further increase in doping ($x>0.04$) when, unlike in the  cuprates, the Fe magnetic moments become even weaker. The resulting phase diagram evidences the presence of a crossover region, where the superconducting and the magnetic order parameters coexist on a nanoscopic range.
\end{abstract}
 %
%

\maketitle

\section{\label{sec:intro}Introduction}
The discovery of high-temperature superconductivity in the iron-based layered compound LaFeAsO$_{1-x}$F$_{x}$ \cite{Kamihara2008} immediately created considerable excitement among condensed matter scientists. Other superconductors belonging to the same RE-1111 family, with RE a rare-earth metal, were discovered successively and superconductivity with a transition temperature of up to 55 K was found when La is substituted by other rare earths as e.g.\ Sm, Ce, Nd, Pr, and Gd.\cite{Chen2008,ChenNAT2008,RenEu2008,RenMat2008,Cheng2008} The new compounds show strong similarities with the high-$T_{c}$ cuprates:\cite{Basov2011} \textit{i}) they have layered crystal structures with alternating REO and FeAs planes, where the iron ions are arranged on a simple square lattice, \textit{ii}) the parent compound is  antiferromagnetically ordered, \textit{iii}) superconductivity emerges upon doping the parent compound with either electrons or holes, a process which suppresses the magnetic order. However, there are also important differences between the two families, among which two are particularly significant: the semi-metallicity of the iron-based parent compound, as opposed to the Mott-insulator character of cuprates, and the moderate degree of electron correlation in the former vs.\ the strong correlation observed in the latter, as from comparative studies of far-infrared reflection.\cite{Basov2009}

The REFeAsO parent compounds generally show a commensurate spin-density wave (SDW) magnetic order characterized by \textit{strongly reduced} Fe momenta, as evidenced by standard neutron scattering studies:\cite{Zhao2008} Fe magnetic moments are comprised between 0.25 and 0.8~$\mu_{\mathrm{B}}$, to be compared with 4 $\mu_{\mathrm{B}}$, the spin-only value for the Fe$^{2+}$ ion. For this very reason, the determination of the temperature dependence of the magnetic order parameter via sensitive local probe techniques, such as muon-spin rotation,\cite{Luetkens2009,Amato2009,Maeter2009,JPCarlo2009} M\"ossbauer spectroscopy,\cite{Klauss2008,McGuire2009} electron spin resonance (ESR),\cite{Alfonsov2011} or nuclear magnetic resonance (NMR)\cite{Bobroff2010} is supposed to provide a higher accuracy than that possible using standard powder neutron scattering.\cite{Lumsden2010} 
Currently, the accepted magnetic moment value for iron at 2 K in the REFeAsO families is 0.63(1) $\mu_{\mathrm{B}}$,\cite{Maeter2009,Qureshi2010} seemingly independent of the rare earth.

The study of magnetism of these systems is crucial: its disappearance often signals the onset of superconductivity. Since the coexistence of the magnetic (M) and superconducting (SC) orders in interspersed nanoscopic domains has been generally associated to unconventional superconductivity, it is of fundamental interest to discover whether the antiferromagnetism persist also in the superconducting phase, or it simply disappears as soon as the SC phase is established. Also the possible role played by the ordering of the rare-earth magnetic moments is not yet clear.

Among the 1111 compounds the case of \cefaof\ is paradigmatic of this situation. It was first investigated via neutron diffraction\cite{Zhao2008,Chi2008}, whose results suggested that both the magnetic and the superconducting order parameters go to zero at the M-SC boundary, thus hinting at the possibility of a quantum critical point (QCP) separating a long-range, anti-ferromagnetically ordered phase from the superconducting phase. This, however, is in contrast with the presence of a nanoscopic coexistence of magnetism and superconductivity, as evidenced in our recent study of \cefaof\ for $x=0.07$.\cite{Sanna2010} 
This apparent discrepancy can be solved by observing that standard neutron diffraction is sensitive only to long-range magnetic order. In addition, also \textcite{JPCarlo2009} have observed an anomalous $\mu$SR relaxation in a similarly doped Ce-compound. To date, though, no systematic studies exist to establish whether the long-range magnetic order becomes short-ranged, or it simply vanishes as the superconductivity appears. A study of the evolution with doping of the magnetic order in the FeAs planes can, therefore, provide us with fundamental hints on the interplay between the  superconductivity and magnetism, i.e.\ whether these two phases are competing or reinforcing each other.

To address the above issues we carried out systematic $\mu$SR investigations in the \cefaof\ system in a large doping range $x$. Due to their local probe character and to the high sensitivity to internal magnetic fields, muons are perfect for unraveling the complex interplay between magnetism and superconductivity in the iron-based superconductors. In the specific case of \cefaof\ we find that as the fluorine content increases the magnetically ordered phase does not disappear, 
even at the highest investigated $x$ concentrations, where it coexists with SC. The features of the 
magnetic order, on the other hand, are found to depend strongly on F doping, with the magnetic coherence range becoming shorter as $x$ increases.
\begin{figure}
\centering
\includegraphics[width=0.43\textwidth]{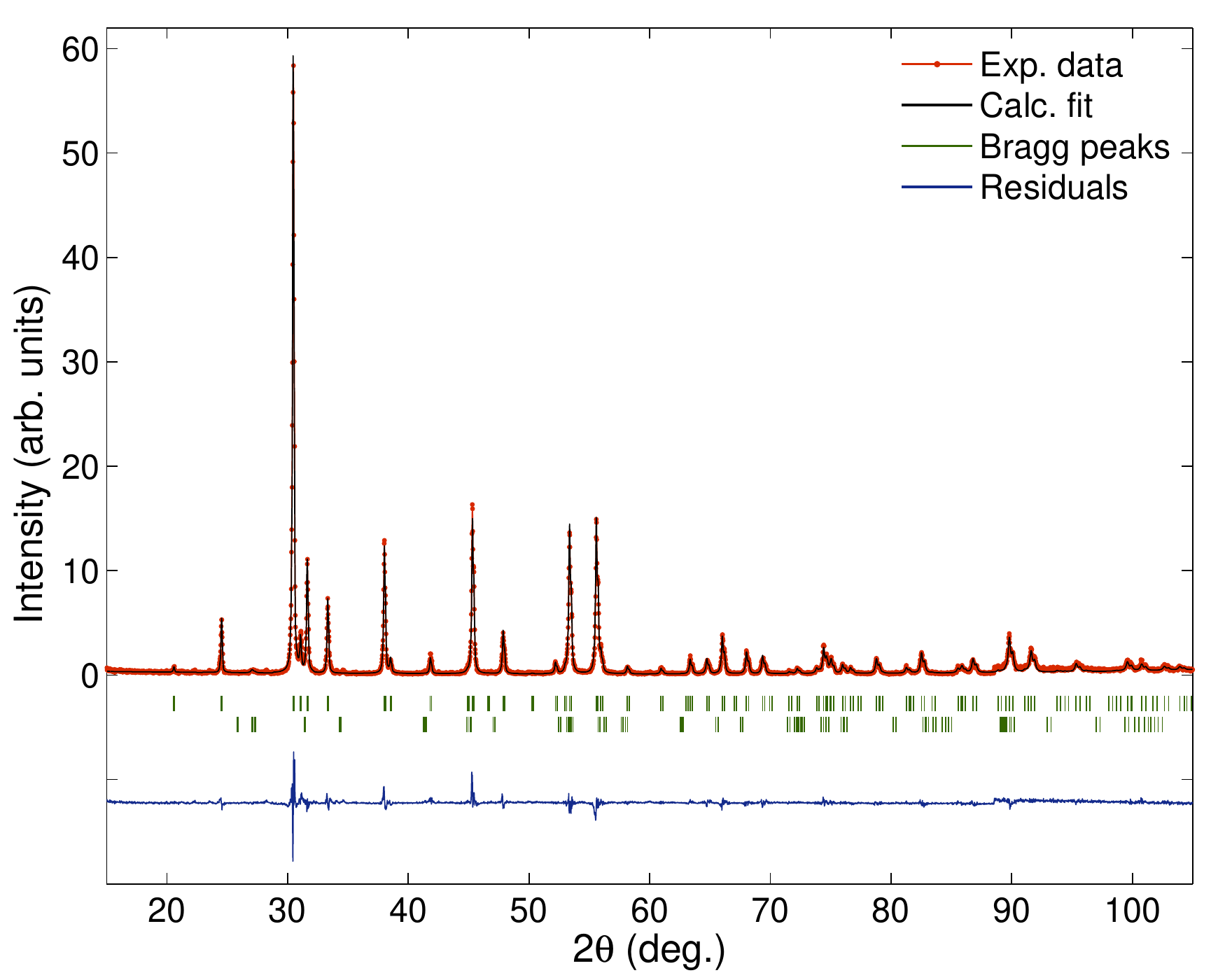} 
\caption{\label{fig:diffrattogramma} (Color online) 
X-ray diffraction pattern for the sample $x=0.04$ and the corresponding Rietveld refinement plot.}
\end{figure}

\section{\label{sec:exp_details}Experimental details and results}
\subsection{\label{ssec:preparation}Sample preparation and morphological and structural characterization}
\begin{figure}
\centering
\includegraphics[width=0.45\textwidth]{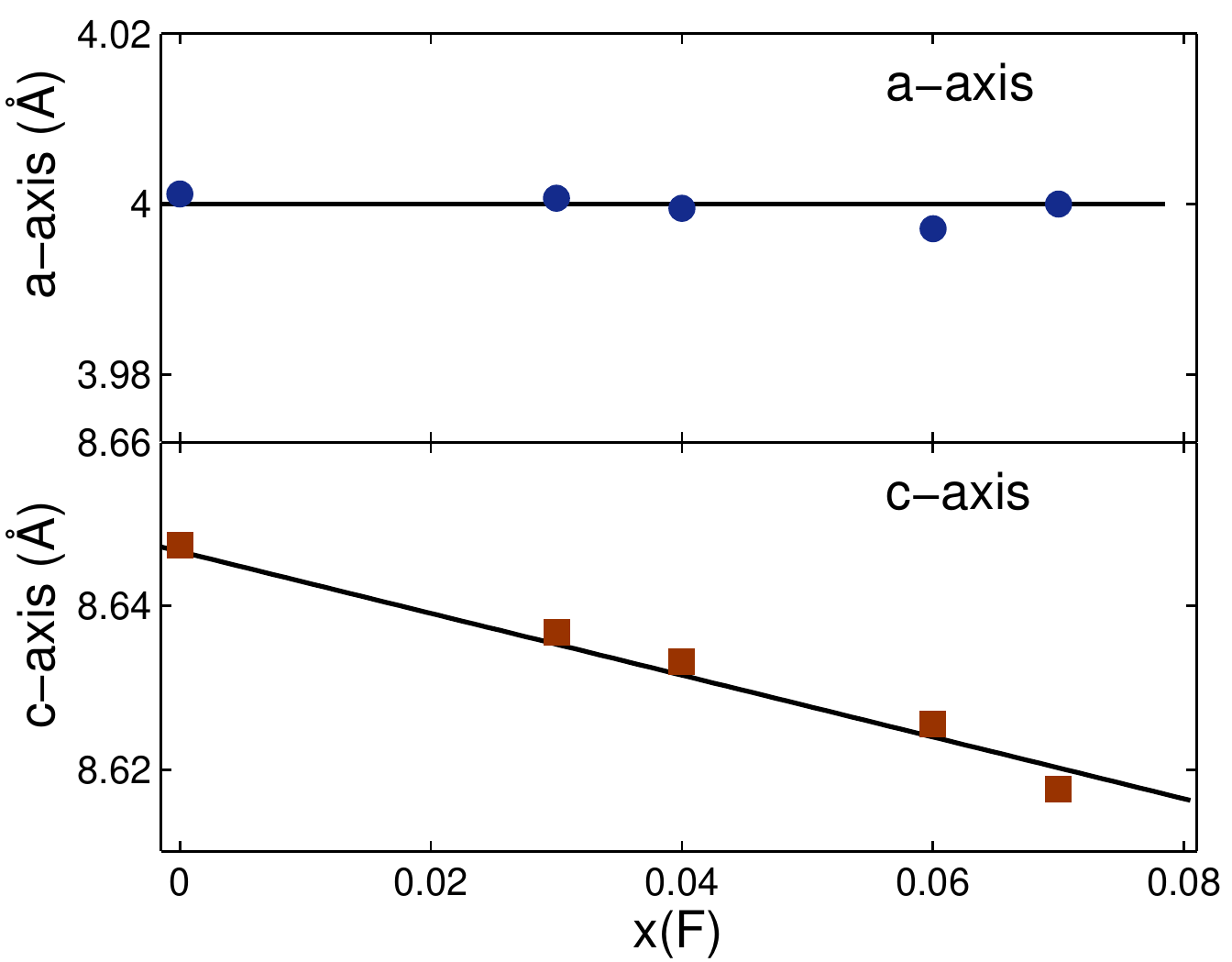}   
\caption{\label{fig:strutturali} (Color online) Evolution of the $a$- and $c$-axis lengths of \cefaof\ as a function of the real fluorine content, $x$(F), at 290 K. Lines are best fits to the experimental data, while the estimated uncertainty $\Delta x/x$ is ca.\ 25\%.}
\end{figure}

A series of \cefaof\ samples, with a real fluorine content ranging from $x=0$ up to $x=0.07$ (see next section), was prepared via solid-state reaction methods by reacting stoichiometric amounts of CeAs, Fe$_{2}$O$_{3}$, FeF$_{2}$, and Fe.
CeAs was obtained by reacting Ce chips and As pieces at 450--500$^{\circ}$C for 4 days.
The raw materials were tho\-rou\-ghly mixed and pressed into pellets, which were then
heated up to ca.\ $1100^{\circ}$C for 50 h.
Further details concerning sample preparation and characterization have been reported elsewhere.\cite{Martinelli2008}
A morphological and a structural characterization were carried out on all the successively investigated samples. The morphological analysis, performed via scanning electron microscopy (SEM), revealed the presence of grains whose dimensions range from 1 to 10 $\mu$m, almost independently from the doping level (see also App.~\ref{ssec:SEM}).
This independence from doping rules out a possible influence of the grain morphology on the magnetic and superconducting properties (e.g., apparent changes in the superconducting fraction, etc.). The structural characterization was carried out using standard powder X-ray diffraction (XRD) analysis. Figure~\ref{fig:diffrattogramma} shows the Rietveld refinement plot for a representative $x=0.04$ sample.
Figure~\ref{fig:strutturali} instead shows the variation of the unit cell parameters as a function
of the real fluorine content. The latter was estimated by means of NMR measurements, as reported below. We note that while the length of the $a$-axis is practically insensitive to F substitution, the $c$-axis' length  decreases linearly with increasing $x$(F), in agreement with data reported in Ref.~\onlinecite{Zhao2008}. As a result, the global effect of O$^{2-}$/F$^{-}$ 
substitution is that of decreasing the volume of the unit cell, a reasonable outcome reflecting the smaller ionic radius of fluorine.
\subsection{\label{ssec:NMR}Determination of fluorine doping via ${}^{19}$F NMR}
The gradual fluorine substitution in a RE-1111 system is known 
to introduce free carriers onto the FeAs layers, hence inducing important
modifications to the material's electronic properties. As such, $x$(F) represents the most natural parameter for describing the phase diagram of F-doped pnictides. Since our aim is the study of the M-SC crossover region, where even tiny variations of \textit{x}(F) could play a major role, the determination of the \textit{real} fluorine content is of crucial importance. To this purpose quantitative fluorine-19 NMR measurements were carried out and the samples classified accordingly.

To obtain a reliable estimate of the $x$(F) values, all the samples were measured using fluorine-free cabling and probehead.
The resonant ${}^{19}$F NMR signal, following a conventional solid-echo sequence ($\pi/2 - \tau - \pi/2 - \tau -$ acquisition),
was acquired in a fixed-field configuration, $\mu_{0}H \simeq 1.4$ T, 
corresponding to $\nu \simeq 56$ MHz for the ${}^{19}$F nuclei.
By recording the NMR signal for different values of the delay $\tau$
we could extra\-po\-la\-te the exponential decay of the integrated echo 
intensity, $I(\tau)$, back to $\tau = 0$. The extra\-po\-la\-tion procedure is 
crucial in providing an unbiased quantitative estimate of the fluorine 
content. In fact, samples with different F doping have, in general, 
different spin-spin relaxation times $T_{2}$, which would potentially 
impair a straightforward intensity comparison.
Since quantitative NMR measurements are notoriously difficult, the 
\textit{total} fluorine content for each sample was evaluated by comparing 
the measured $I(0)$ values with that of an SmOF reference compound.\cite{Sanna2010}
In addition, we recall that NMR alone is unable to distinguish the fluorine signal 
coming from primary vs.\ secondary phases. Therefore, to obtain the \textit{true} $x$(F) 
values, the measured NMR intensity data were finally corrected to take into 
account the possible presence of CeOF impurities. The relative amount of the latter 
was accurately quantified by means of Rietveld refinement of X-ray powder diffraction patterns. 
We find that, whenever present, the spurious CeOF phase never exceeds 2\% vol.
This analysis reveals that the real F content is systematically lower than the nominal 
one but, nevertheless, the use of NMR labeling is superior to the simple use of the 
nominal doping. Therefore, hereafter we use the former as a label for the different samples. 
Possible errors would uniformly affect the investigated samples, implying a rescaling of 
the final phase diagram, but cannot give rise to distortions or inversions.

\subsection{\label{ssec:transport}Transport measurements}
\begin{figure}[bht]
\centering
\includegraphics[width=0.45\textwidth]{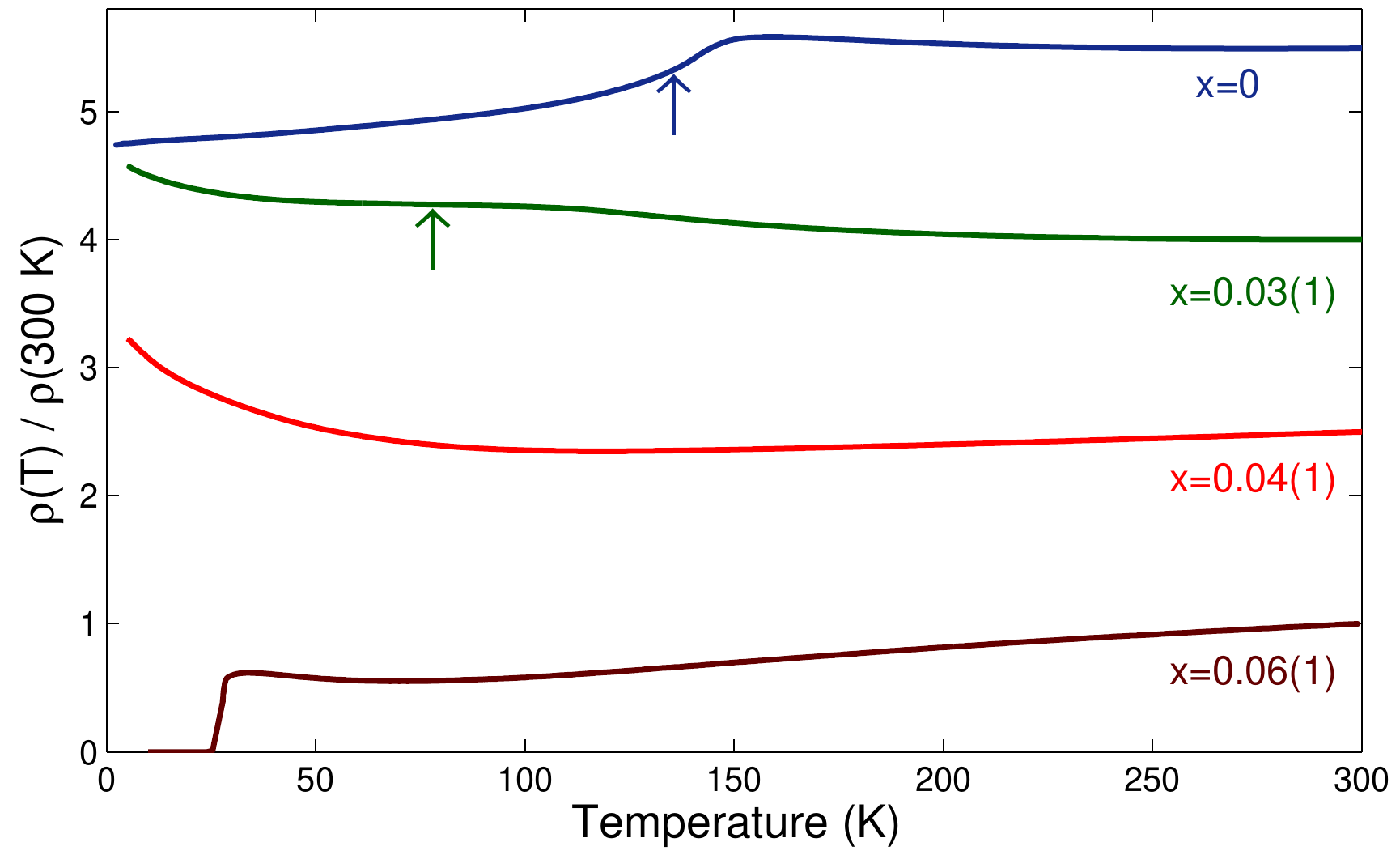}    
\caption{\label{fig:trasporto} (Color online) Normalized resistivity vs.\ temperature for a selection of \cefaof\ samples. For a better visibility the curves have been shifted against each other by 1.5 units. Arrows indicate the maxima of $\mathrm{d}\rho/\mathrm{d}T$.}
\end{figure}

The resistivity of the \cefaof\ samples was measured using a standard four-point method. The temperature dependence of $\rho(T)$ for selected F doping values is shown in Fig.~\ref{fig:trasporto}. Upon cooling, the undoped sample presents the typical transport features of iron-based oxypnictides:  \textit{i}) a low-temperature resistivity in the m$\Omega\,$cm range; \textit{ii}) a broad maximum, followed by a drop of $\rho(T)$, with an inflection point defined as the maximum of the first derivative, $\mathrm{d}\rho/\mathrm{d}T$ (arrows in Fig.~\ref{fig:trasporto}).
The presence of a maximum in the first derivative of resistivity has been observed also in the 1111 systems containing La,\cite{Klauss2008,McGuire2008} Pr\cite{Kimber2008} and Sm,\cite{Tropeano2009} and 
has been generally attributed to a spin-density wave (SDW) transition. As the fluorine content increases, the maxima of $\rho(T)$ and $\mathrm{d}\rho/\mathrm{d}T$ both shift towards lower temperatures, become broader and eventually disappear for $x=0.04$, in full agreement with existing experimental data on the Ce-1111 family.\cite{Chen2008}

\subsection{\label{ssec:magnetization}Magnetization measurements}
\begin{figure}
\centering
\includegraphics[width=0.45\textwidth]{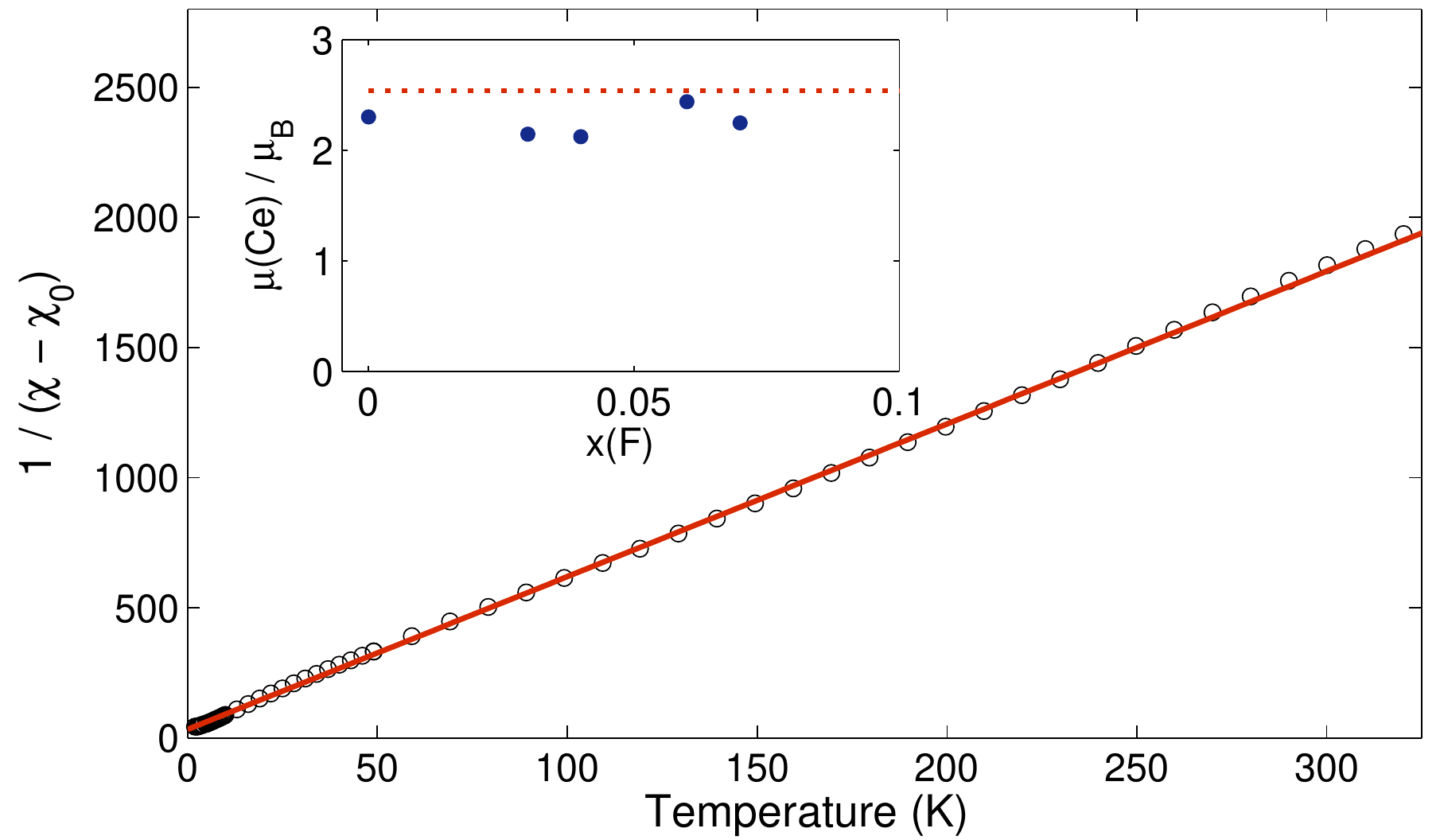}       
\caption{\label{fig:mag1} (Color online) Linear fit of $1/(\chi - \chi_0$) for the case of the undoped sample ($x=0$). Inset: Ce free-ion magnetic moment for all the samples under test, as derived from Curie-Weiss fits of the dc magnetization data. The dotted line shows the theoretically expected value 
for free Ce$^{3+}$ ions (2.54 $\mu_{\mathrm{B}}$).}
\end{figure}

\begin{figure}
\centering
\includegraphics[width=0.45\textwidth]{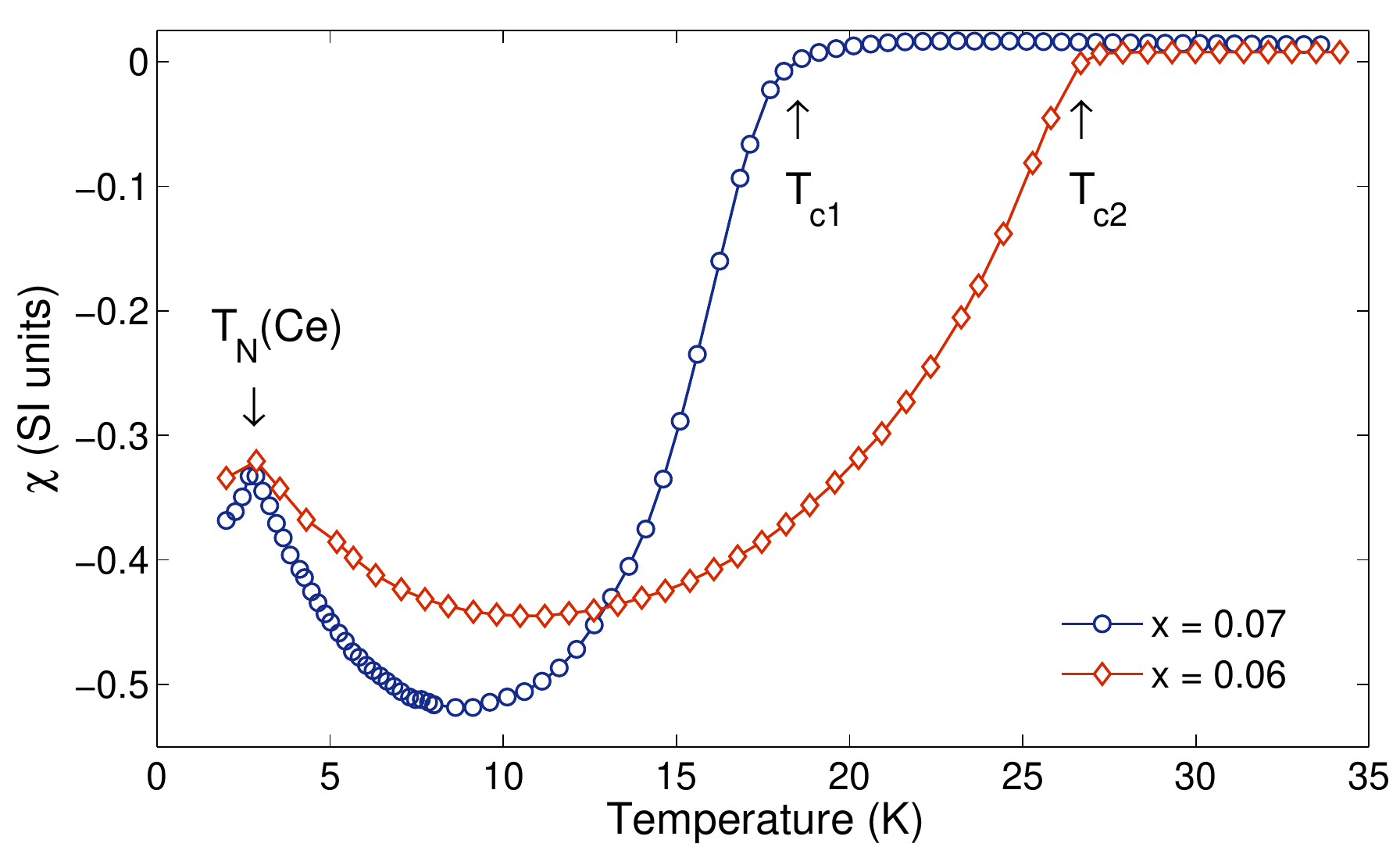}      
\caption{\label{fig:magSC} (Color online) ZFC magnetization data for the superconducting samples at $\mu_0 H = 0.5$~mT. Arrows denote the FeAs SC transitions and the Ce magnetic ordering temperatures, respectively.}
\end{figure}

DC magnetization measurements were performed by means of a superconducting quantum interference device (SQUID) magnetometer (Quantum Design) on all the tested samples. 
Both magnetization vs.\ temperature, from 2 up to 300 K at $\mu_0 H = 3$~T, in zero-field cooling (ZFC) 
and in field cooling (FC) conditions, as well as magnetization measurements vs.\ applied field, $m(H)$, at selected temperatures were carried out. The experimental results can be summarized as follows:\\[1.5mm]
\textit{i}) The quantity of dilute ferromagnetic impurities, if any, is irrelevantly small. This is evinced by the linear (i.e.\ purely paramagnetic) behavior of $m(H)$ at both low and high temperatures (not shown).\\[1.5mm]
\textit{ii}) Ce ions mostly retain their free-ion magnetic moment value. This result arises from numerical fits of $\chi(T)$ data using the Curie-Weiss law, $\chi(T)=C/(T-\theta)+\chi_0$, with $\chi_0$ the temperature-independent susceptibility, $C$ the Curie constant, and $\theta$ the Curie-Weiss temperature. First, we determine $\chi_0$ by considering only the high-temperature regime. Successively, we perform a linear fit of $1/(\chi(T)-\chi_0)$, as shown in Fig.~\ref{fig:mag1} for a typical case, $x=0$. From the resulting Curie constant one can determine the Ce magnetic moment (in the free-ion approximation), which is plotted against the F content in the inset of Fig.~\ref{fig:mag1}.
This value is very close to that expected for free Ce$^{3+}$ ions (2.54 $\mu_{\mathrm{B}}$) and in perfect agreement with previous data.\cite{Chen2008,Zhao2008,CeION_1,CeION_2}\\[1.5mm]
\textit{iii}) An increase in F doping up to $x = 0.07$ depresses slightly the cerium antiferromagnetic (AF) ordering temperature, $T_{\mathrm{N}}$(Ce). 
We should here caution the reader that the Ce AF transition is somehow affected by applied fields of moderate intensity: a magnetic field of $\sim 4$~T 
can lower an otherwise ``regular'' $T_{\mathrm{N}}$(Ce) by more than 2 K, in agreement 
with data reported in Ref.~\onlinecite{Chen2008,Riggs2009}.
All $T_{\mathrm{N}}$(Ce) values, measured in similar low-field conditions, are reported in Table~\ref{tab:table}.\\[1.5mm]
\textit{iv}) Samples with $x \gtrsim 0.06$ 
display a clear transition to the superconducting state, as shown by the low-temperature ZFC magnetization data for 
$x=0.06$ and 0.07 reported in Fig.~\ref{fig:magSC}. A precise determination of the superconducting 
fraction from the magnetization data, though, is difficult: at lower F-doping values, the field penetration depth increases 
considerably and becomes comparable to the grain size (1--10 $\mu$m), thus effectively reducing the shielding volume 
within each grain. Nevertheless, TF-$\mu$SR data clearly demonstrate that all these samples show bulk superconductivity, 
as reported in detail for the $x=0.07$ case in Ref.~\onlinecite{Sanna2010}.
\begin{table*}[thb]
\centering
\renewcommand{\arraystretch}{1.2}
\caption{\label{tab:table}Magnetic and superconducting properties of \cefaof\ 
for different values of F doping.}
\begin{ruledtabular}
\begin{tabular}{p{1mm}lcccccc}
\vspace{2mm}
& \lower 0.7mm \hbox{$x$(F)}
& \lower 0.7mm \hbox{$T_{\mathrm{N}}$ (K)}
& \lower 0.7mm \hbox{$\Delta T_{\mathrm{N}}$ (K)}
& \lower 0.7mm \hbox{$T_c$ (K)}
& \lower 0.7mm \hbox{Magnetic frac.\ (\%)}
& \lower 0.7mm \hbox{$T_{\mathrm{N}}$(Ce) (K)}\\[2pt]
\hline
&0        &150(2)   &10(1)  &--       & 97(2)  &4.0(3) \\
&0.03(1)  &106.0(9) &9(1)   &--       & 93(1)  &3.5(3) \\
&0.04(1)  &37.9(9)  &9(1)   &--       & 100(8) &2.78(3)\\
&0.06(1)  &28.0(7)  &11(1)  &26.5(5)  &100(3)  &2.7(3) \\
&0.07(1)  &16.5(3)    &9.8(0.4)   &18.3(5)  &100(5)  &2.8(3) \\
\end{tabular}
\end{ruledtabular}
\end{table*}
\subsection{\label{ssec:ZFmuSR}ZF-$\mu$SR measurements}
The muon-spin relaxation measurements were carried out at the GPS instrument ($\pi$M3 beam line) of the Paul Scherrer Institut, Villigen, Switzerland. $\mu$SR experiments consist in implanting 100\% spin-polarized muons in a sample and successively detecting the relevant decay positrons. Implanted muons thermalize almost instantaneously at interstitial sites, where they act as sensitive probes, precessing in the local magnetic field $B_{\mu}$ with a frequency $f_{\mu} = \gamma/(2\pi)\cdot B_{\mu}$, with $\gamma/(2\pi) = 135.53$ MHz/T the muon gyromagnetic ratio. Both zero-field (ZF) and longitudinal field (LF) $\mu$SR experiments were performed. Due to the absence of applied fields, ZF-$\mu$SR represents the best technique for investigating the spontaneous magnetism and its evolution with fluorine doping. LF-$\mu$SR measurements, instead, were used to single out the dynamic or static character of the magnetic order, as detailed in App.~\ref{ssec:LFmuSR}.

\begin{figure}[t]
\centering
\includegraphics[width=0.46\textwidth]{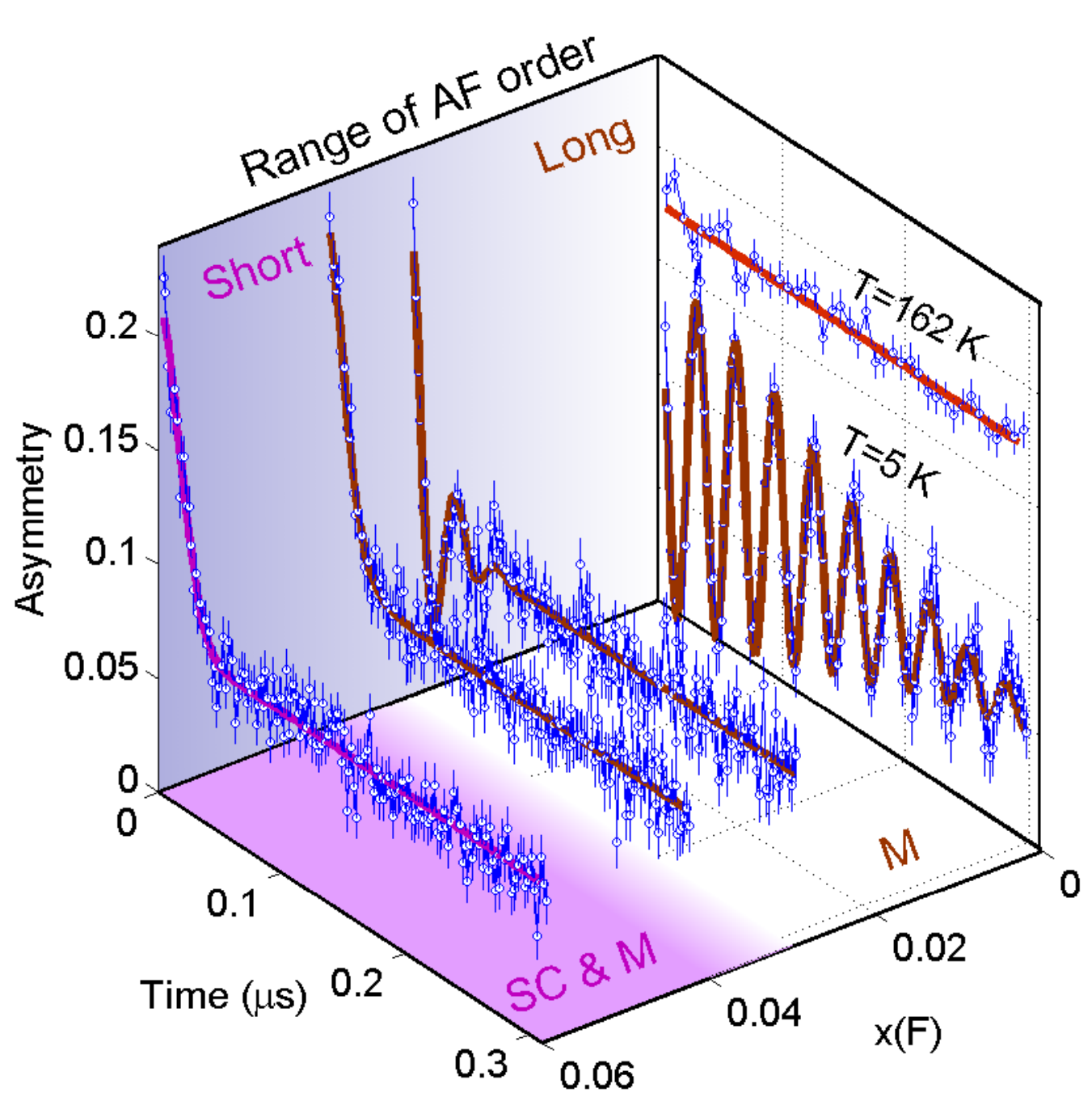}         
\caption{\label{fig:LongShort} (Color online) Zero-field, time-domain $\mu$SR data of \cefaof\ for selected fluorine doping values $x$, recorded at $T = 5$ K. The superconductivity appears for $x \gtrsim 0.06$, whereas the antiferromagnetic order is always present. The range of the AF order, however, changes from long to short, as shown by the absence of oscillations in samples with a high F content.}
\end{figure}

Figure~\ref{fig:LongShort} shows the low-temperature ZF asymmetry (i.e.\ the  muon-spin precession signal), $A(t)$, for the tested \cefaof\ samples. As a comparison, the high-temperature asymmetry is also shown for the undoped $x=0$ case. At high temperatures, i.e.\ above the N\'eel temperature $T_{\mathrm{N}}$, the asymmetry signal is practically flat, with no oscillations and with a very small decay of the initial polarization. This behavior is typical of paramagnetic materials, where there are no significant internal magnetic fields, except for those due to the tiny randomly-oriented nuclear moments, which account for the exiguous decay of asymmetry. 
%

\begin{figure}
\centering
\includegraphics[width=0.45\textwidth]{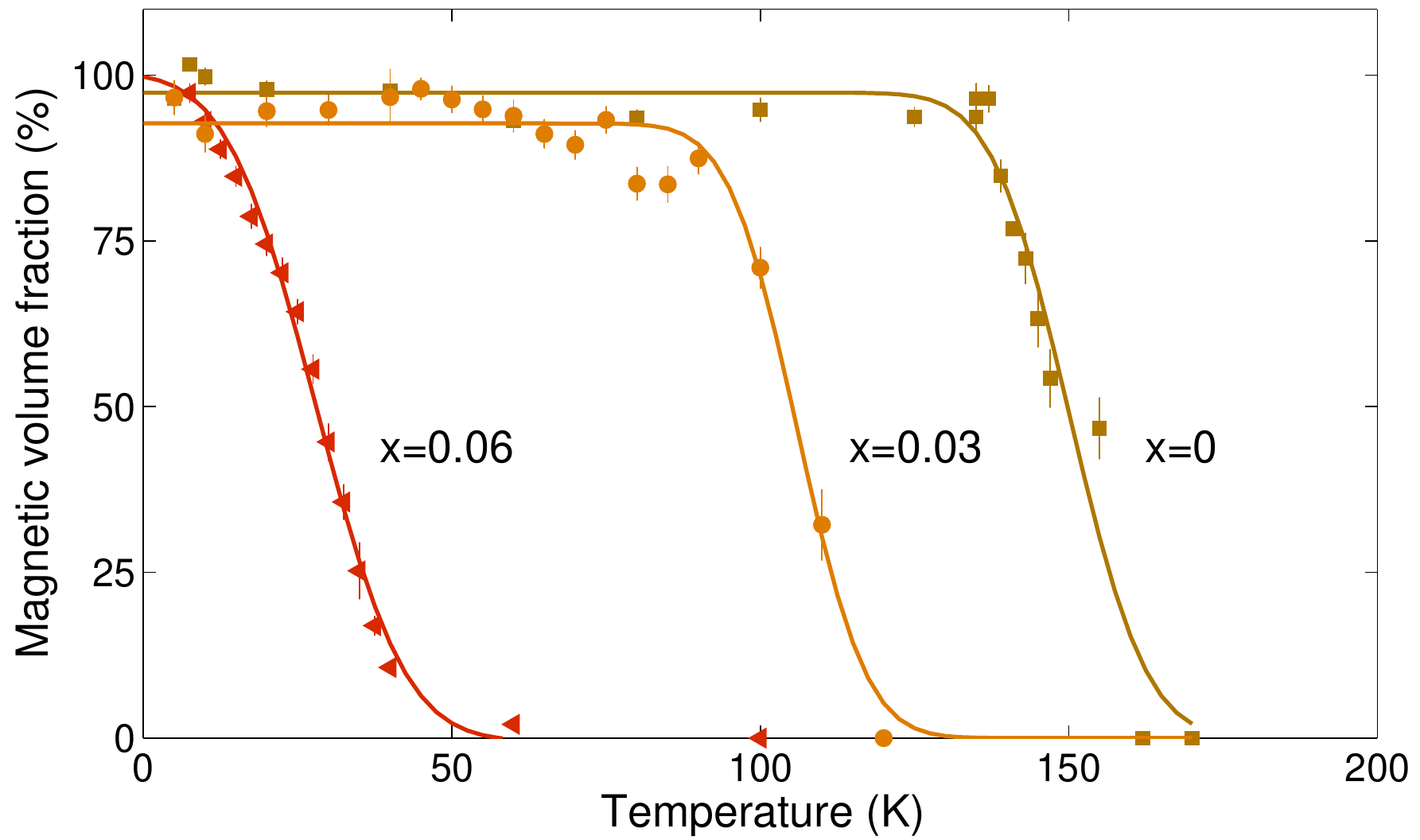}     
\caption{\label{fig:volMag} (Color online) Magnetic volume fraction vs.\ temperature as extracted from the longitudinal component of $\mu$SR data and fitted with an $\mathrm{erf}(T)$ function (Eq.~\ref{eq:mag_vol_frac}) for a selection of samples ($x=0$, 0.03, 0.06). The relevant fit values, including those of other samples, are reported in  Table~\ref{tab:table}.}
\end{figure}
However, once the temperature is lowered below $T_{\mathrm{N}}$, dramatic differences develop, reflecting the emergence of a spontaneous magnetic order. The new ordered phase seems to have features which depend strongly on $x$: 
for a low F content we observe well-defined asymmetry oscillations which, as $x$(F) increases, change quickly to highly damped ones.
The damping for $x \ge 0.04$ becomes so high that the oscillatory behavior disappears altogether, to be replaced by a fast decaying signal. We recall that asymmetry oscillations indicate the presence of a uniform magnetic field at the muon sites, while a strong decay of the asymmetry arises whenever there is a wide distribution of fields. Hence, the case $x=0$ is compatible with a \emph{long-range} (static) magnetic order. On the other hand, a strong $\mu$SR signal decay (for $0.04\lesssim x \lesssim 0.07$) is compatible with dephasing (i.e.\ incoherent muon-spin precession) due to a distribution of internal fields, which can be attributed to a \emph{short-range} magnetic order. 
%
In Fig.~\ref{fig:LongShort} the evolution with doping of the spin polarization of the muon ensemble is followed against the fluorine content.

To achieve a better understanding from the above results, the various ZF-$\mu$SR time-domain data were fitted using the function:
\begin{equation}
\label{eq:osc}
\frac{A^{\mathrm{ZF}}(t)}{A_{\mathrm{tot}}^{\mathrm{ZF}}(0)}= \left[ a^{A}_{\perp} e^{-\frac{\sigma_A^2 t^2}{2}} f (2 \pi \gamma B_{\mu} t) \!
+ a^{B}_{\perp} e^{-\frac{\sigma_B^2 t^2}{2}}\right]+ a_{\parallel}e^{-\lambda t}.
\end{equation}
Here $a^{A,B}_{\perp}$ are two transverse components, while $a_{\parallel}$ represents a longitudinal component, each with respective decay coefficients $\sigma_{A,B}$ and $\lambda$. The choice of two transverse decay components follows closely the calculations presented in Ref.~\onlinecite{Maeter2009}, according to which, for the undoped CeFeAsO case (but largely valid also in presence of F doping), two distinct muon implantation sites are expected. The most populated site, named A, is located next to the FeAs planes, while a second site, named B, is close to the oxygen atoms in the CeO planes. As suggested, muons implanted in A are sensitive to a magnetic field that consists of two contributions: the molecular field generated by the Fe sublattice and the field arising from the Ce polarization, the latter being induced by the Fe magnetism (via an exchange coupling $J_{\mathrm{Fe-Ce}} \sim 43$ T/$\mu_{\mathrm{B}}$).\cite{Maeter2009}
Muons implanted in B are sensitive mostly to this second contribution. Since the latter site is statistically the least populated one (accounting for only $\sim 15\%$ of the implanted muons, as from Eq.~\ref{eq:osc}), its contribution to the total asymmetry does not permit to distinguish a second precession frequency and therefore it is taken into account by a single exponential decay ($a^{B}_{\perp}$ ). The corresponding longitudinal components, nevertheless, share similar decay rates, which allows us to merge them in a single term, $a_{\parallel}$. 

The nature of the oscillating term $f(t)$ depends on the F content: for $x=0$ (and other low $x$ values) $A(t)$ could be fitted using $f(t)=\cos(2\pi\gamma B_{\mu} t)$; on the other hand, for $x=0.03$ the best fit was obtained with $f(t)=J_0(2\pi\gamma B_{\mu} t)$ (here $J_0$ is the zeroth order Bessel function). The cosine term is the hallmark of \textit{commensurate} long-range magnetic order, while the presence of a Bessel function is generally attributed to \textit{incommensurate} long-range ordered systems.\cite{SAVICI} Finally, $f(t)=1$ for all those samples where no coherent oscillations could be detected ($x \gtrsim 0.04$). As a result, the observed static AF order seems to be commensurate for $x < 0.03$, incommensurate for $x \sim 0.03$ and fully disordered for $0.04 \lesssim x \lesssim 0.07$.
\begin{figure}
\centering
\includegraphics[width=0.45\textwidth]{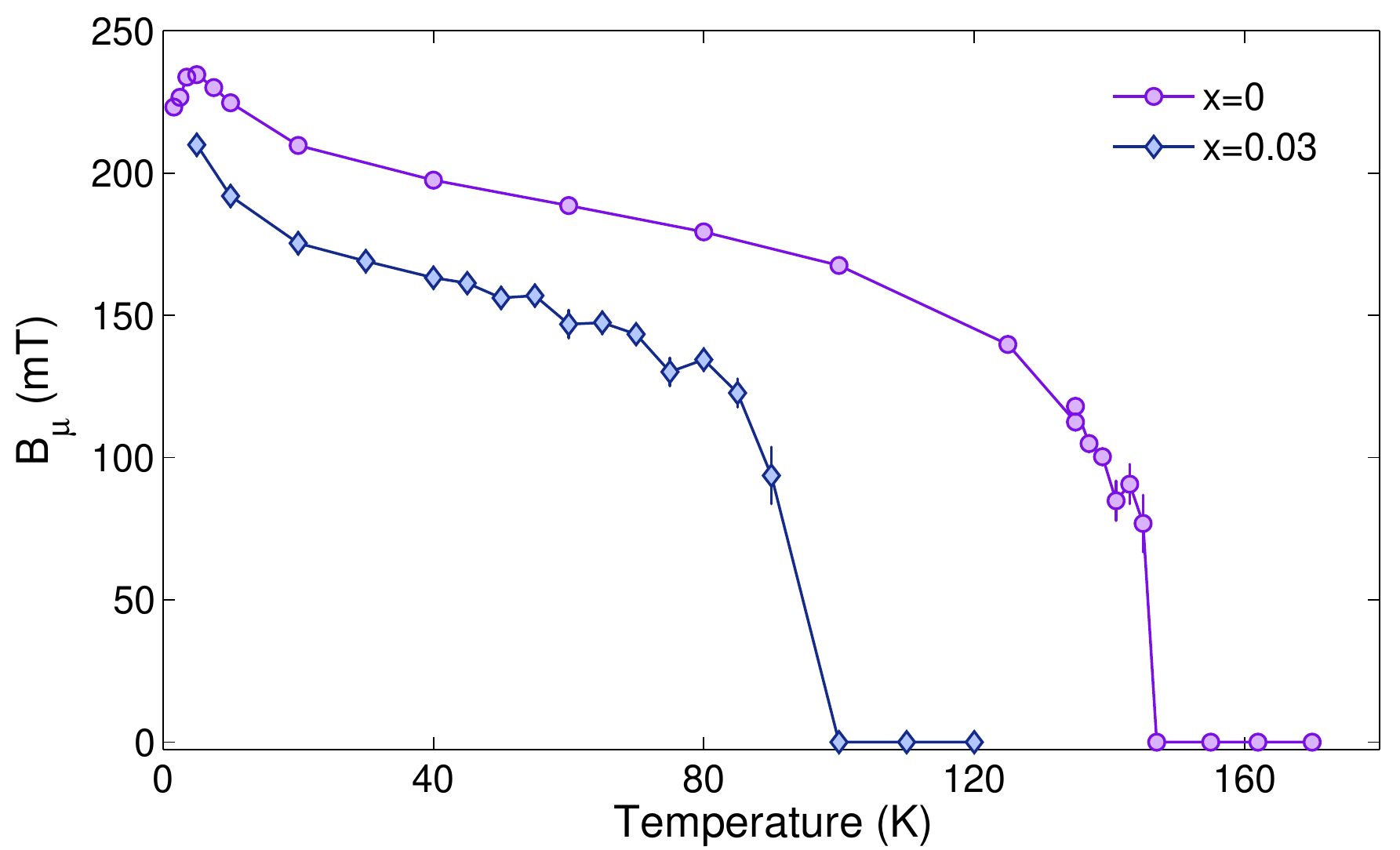}        
\caption{\label{fig:Bmu}(Color online) Internal magnetic field $B_\mu$ as probed by ZF-$\mu$SR at the so-called A-sites for $x=0$ and $x=0.03$.}
\end{figure}

Let us now determine the magnetic volume fraction from the $\mu$SR data. Since all the considered samples were available as pressed powder pellets, we can safely assume a 
randomly oriented internal field model: in an ideal case, where the whole sample shows static antiferromagnetic order, on average, $\nicefrac{1}{3}$ of the muons experience a field parallel to their initial spin direction, and hence do not precess (longitudinal component), while the remaining  $\nicefrac{2}{3}$ will precess (transverse component). This is exactly what we find for all the measured samples. From the temperature dependence of the longitudinal component one can follow the evolution of the volume fraction, $V_M$, of the magnetically ordered phase, $V_M(T) = \frac{3}{2} \left(1-a_{\parallel}\right) \cdot 100\%$. The resulting $V_M(T)$ values for a selection of representative samples are plotted in Fig.~\ref{fig:volMag}. It is worth noticing that \emph{all} samples with $x \lesssim 0.07$ become fully magnetic at low temperature ($V_M = 100\%$).
To determine the average N\'eel temperature, the corresponding transition width and the magnetically ordered fraction, the obtained $V_M(T)$ data were fitted using the following phenomenological function: 
\begin{equation}
\label{eq:mag_vol_frac}               
V_M(T) = \frac{1}{2} \left[ 1 -\mathrm{erf} \left(\frac{T-T_{\mathrm{N}}}{\sqrt{2}\Delta}\right)\right].
\end{equation}
The fit results, summarized in Table~\ref{tab:table}, as well as in  Fig.~\ref{fig:volMag} for selected cases, clearly show that as the fluorine content $x$(F) increases there is both a gradual decrease of $T_{\mathrm{N}}$ and a progressive broadening of the transition.

\begin{figure}
\centering
\includegraphics[width=0.485\textwidth]{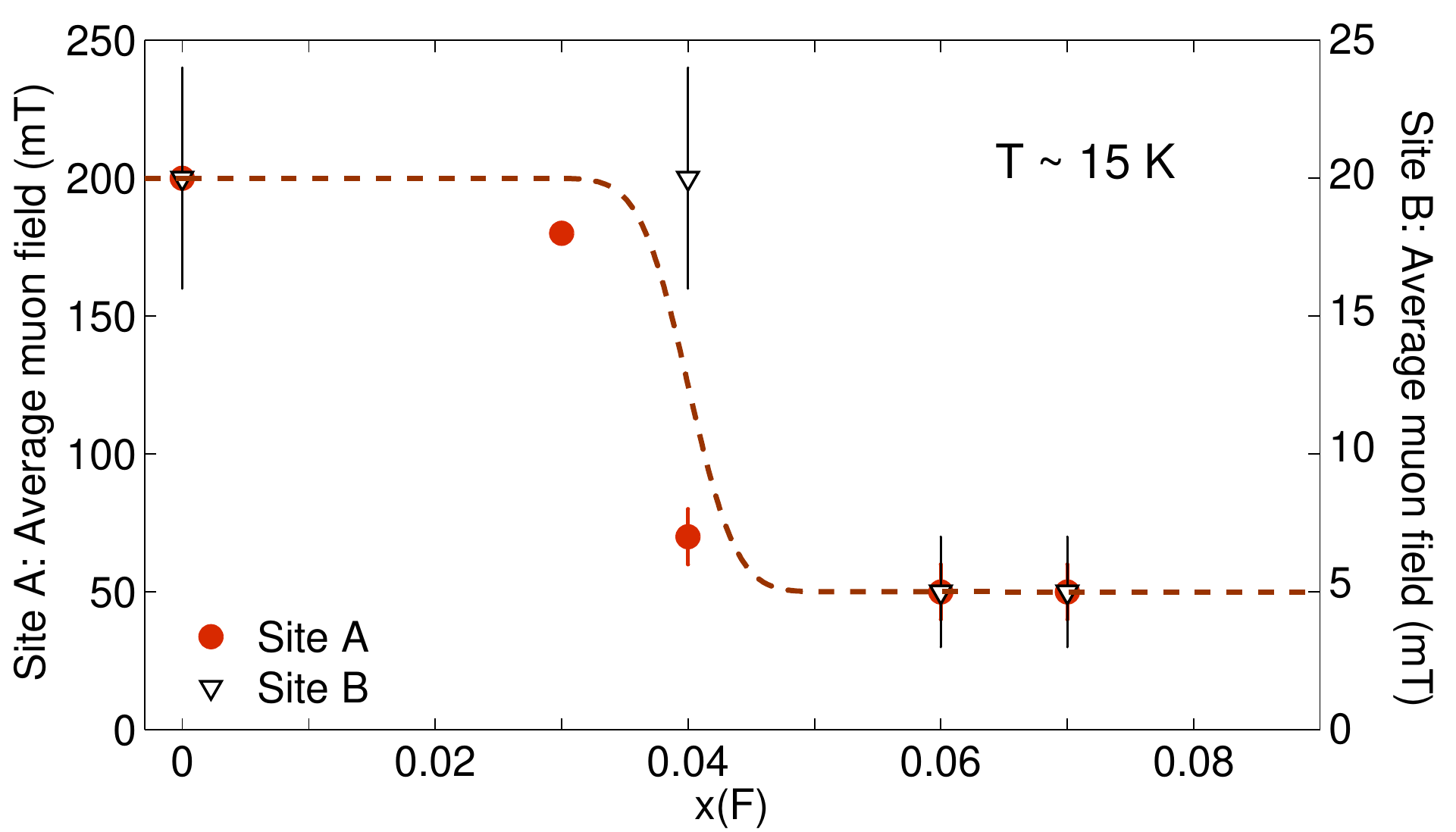}    
\caption{\label{fig:largRIGA}(Color online) Dependence on fluorine doping of the average 
magnetic field value $\left< B_{\mu} \right>$ (for $x<0.4$) and field width $\left<\Delta B_{\mu}^2 \right>^{1/2}$ (for $x \ge 0.4$), as measured by muons stopping in sites A and B. 
The dashed line provides a guide to the eye, emphasizing 
the sharp field drop at $x \sim 0.4$, in concomitance with the onset of the short-range magnetic order.}
\end{figure}
The internal field $B_{\mu}$, as resulting from fits of the asymmetry data with Eq.~\ref{eq:osc}, is plotted in Fig.~\ref{fig:Bmu} for the case $x=0$ and 0.03. In the undoped sample $B_{\mu}$ corresponds to values of the order of $\sim 200$~mT (27~MHz), which are typical for the oxypnictides\cite{Amato2009} and in full agreement with those reported in Ref.~\onlinecite{Maeter2009}. As for the temperature dependence, $B_{\mu}$ shows a characteristic low-$T$ increase (instead of a saturation), which is peculiar of CeFeAsO\cite{Maeter2009} and is due to the AF ordering of Ce$^{3+}$ ions. As the fluorine content increases the average Gaussian field value $\left< B_{\mu} \right>$ reduces and its width broadens. At high F doping, no more coherent oscillations are present in $A(t)$. In these cases the internal field can be described as a broadened distribution of fields whose values range from zero up to $\sigma/\gamma\! = \!(\overline{B_i}^2- \overline{B_i^2})^{1/2}$.\cite{Kadono2004}

Figure~\ref{fig:largRIGA} displays the behavior of the internal field for  $T_{\mathrm{N}}\mathrm{(Ce)} < T \ll T_{\mathrm{N}}$ as a function of F content for both muon sites A and B. Since the internal field is proportional  (via the dipolar interaction) to the staggered magnetization due to the Fe ordered moments, plus a less relevant contribution from the Ce polarized sublattice,\cite{Maeter2009} Fig.~\ref{fig:largRIGA} shows that the Fe magnetic moment is progressively reduced as the doping content increases. This is a general feature of the iron superconductors in marked contrast with the cuprates. 
In fact, unlike in the former, in the cuprates the low-temperature staggered magnetization always recovers to the value of the undoped parent compound. This recovery occurs throughout the doping range corresponding to a magnetically ordered phase,\cite{Borsa1995} either long- or short-ranged, and irrespective of its possible coexistence with superconductivity.\cite{Coneri2010,Sanna2010b}  
This feature is a consequence of the Mott-Hubbard character of the cuprates and the related spin freezing, both absent in the iron pnictides.

\section{\label{sec:discussion}Discussion}
\begin{figure}
\includegraphics[width=0.45\textwidth]{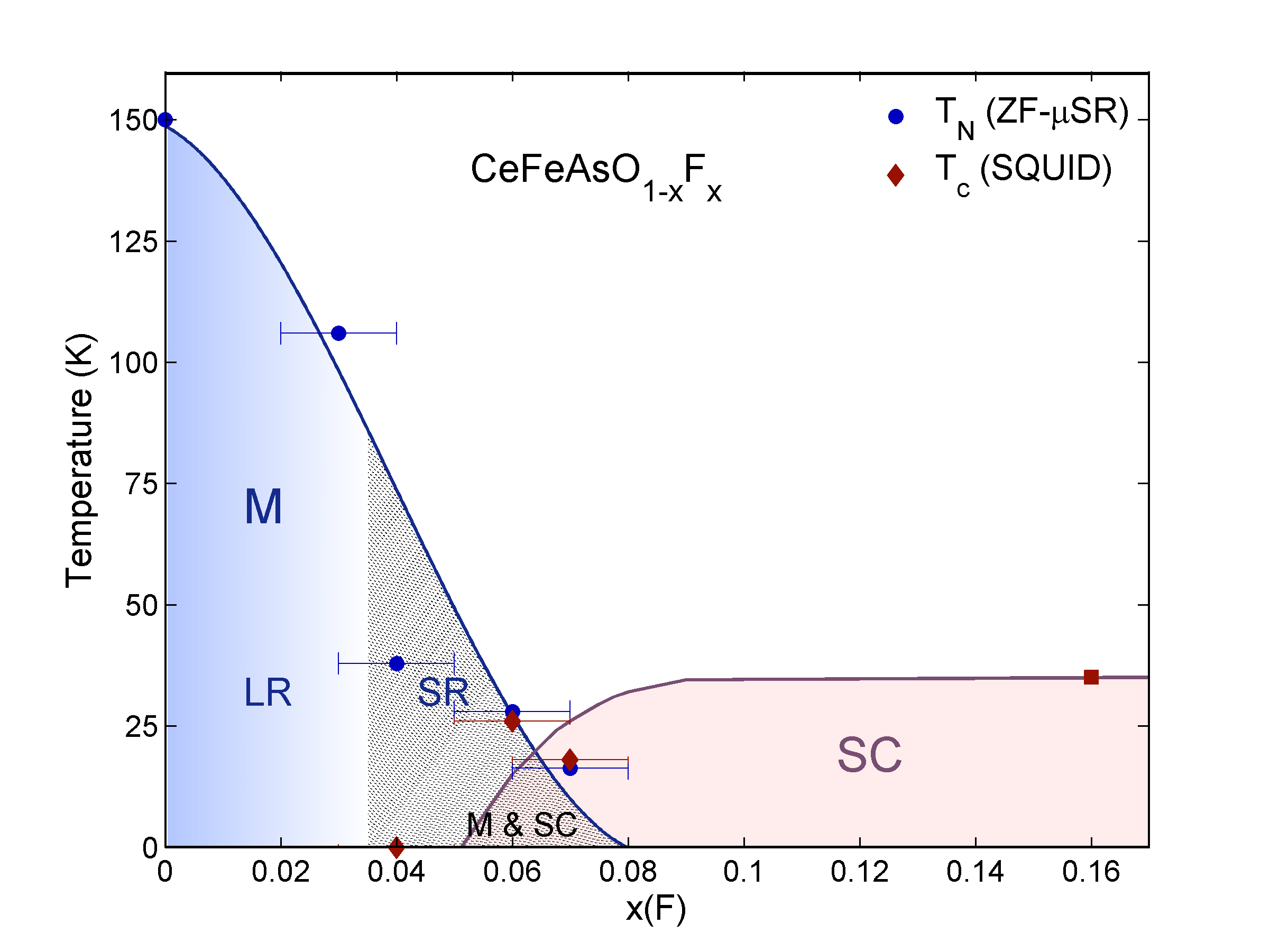}   
\caption{\label{fig:phase_diag} (Color online) Phase diagram of \cefaof\ as determined from the current $\mu$SR measurements. 
M and SC coexist in the violet region, while the hatched area indicates the presence of a short-range (SR) magnetic order. The point at $x = 0.16$ was taken from Ref.~\onlinecite{Zhao2008}. See text for details.}
\end{figure}

The study of doping effects on both the magnetic and the superconducting properties, as well as the presence of a possible M-SC crossover region in the Ce-1111 family compounds, are as interesting as important to our comprehension of the new iron arsenide compounds. To this aim, we have performed a full investigation of structural, transport and magnetic properties on a series of samples with $x$(F) ranging from 0 up to 0.07. By collecting the results from all the presented investigations we obtain a coherent physical picture that provides us with important hints on the evolution with doping of the magnetic order in the FeAs planes and on the influence of the latter on the developing superconducting order. In particular, we highlight  the following key results:\\[1.5mm] 
\textit{i}) The Ce$^{3+}$ AF ordering temperature decreases gradually as the doping $x$(F) increases. This behavior most likely suggests a correlation between the magnetic order of iron and that of the cerium ions.\\[2mm]
\textit{ii}) In the undoped compound, a long-range commensurate AF order sets in for temperatures below $T_{\mathrm{N}}$. This long-range magnetic order is evidenced as coherent oscillations of the muon polarization at low temperatures. Once a small percentage of fluorine is substituted to oxygen, these oscillations become highly damped. In particular, for the $x=0.03$ case a zeroth-order Bessel function seems to provide a better fit to $A(t)$ than a harmonic (cosine) function, thus signaling the onset of an incommensurate AF order of iron ions.\cite{SAVICI,UEMURA} At higher F doping, the shrinking dimensions of the magnetically ordered domains imply a faster dephasing and depolarization of the implanted muons. This marked depolarization is reflected in such a high asymmetry damping 
that it prevents the detection of any coherent oscillations.\\[2mm]
\textit{iii}) Our $\mu$SR results can be summarized in the revised phase diagram shown in 
Fig.~\ref{fig:phase_diag}. Here we report the ordering temperatures, $T_{\mathrm{N}}$, 
as obtained by fits of $a_{\parallel}(T)$ asymmetry using Eq.~(\ref{eq:mag_vol_frac}), and 
$T_c$, in the superconducting phase, as determined by dc magnetization measurements.
Our data are in a fairly good agreement with those of Ref.~\onlinecite{Zhao2008} for $x<0.04$, where some long-range magnetic order still persists and, therefore, is detectable also by neutrons. However, for higher F concentrations there is a clear discrepancy, most likely due to the fact that ordinary powder diffraction methods using thermal neutrons are not sensitive to short-range order. In fact, from $x \geq 0.04$ and up to $x = 0.07$, muons are able to evidence the presence of a \emph{short-range static magnetic order} (hatched area in Fig.~\ref{fig:phase_diag}).
In the narrow region, extending from $x \sim 0.05$ to 0.07, this magnetic order coexists with superconductivity on a 
nm-range scale, as detailed in Ref.~\onlinecite{Sanna2010} (thistle-colored area in Fig.~\ref{fig:phase_diag}), then magnetism is expected to disappear above $x=0.07$. 
The narrow M-SC coexisting region clearly rules out the presence of a quantum 
critical point, i.e.\ the existence of a single critical F-doping separating the two phases. 
Consequently, our results are in good agreement with a recently suggested Ce-1111 
phase diagram,\cite{Wang2011} where the long-range AF and SC phases do not overlap 
but are separated by an intermediate phase, which we identify with the short-range AF phase.

\section{Conclusion}
The evolution of the magnetic order vs.\ F doping in \cefaof\ was mapped via muon-spin spectroscopy. The experimental data and the successive analysis confirm the coexistence of two magnetic-ion subsystems, one related to Ce$^{3+}$ and the other to Fe$^{2+}$. The cerium antiferromagnetism, occurring at relatively low temperatures ($< 4$~K), seems to be correlated with the magnetic order taking place in the FeAs planes, as evidenced by the slight drop of $T_{\mathrm{N}}$(Ce) observed when the Fe magnetic order disappears.
The FeAs magnetic order,  a hallmark also for the Ce-1111 family, is strongly affected by an increase in F content: \textit{i}) its N\'eel temperature decreases sharply; \textit{ii}) the AF magnetic order evolves from \emph{long-} to \emph{short-}ranged; \textit{iii}) superconductivity appears when the Fe$^{2+}$ magnetic moment is significantly reduced with respect to its value in the undoped case. These findings seriously question the presence of a quantum critical point in Ce-1111 and, together with previous results on Nd-\cite{JPCarlo2009}, Sm-\cite{Sanna2009} and Gd-1111\cite{Alfonsov2011} compounds, most likely suggest that the phase diagram of the 1111 family is RE-independent and consists of a narrow M-SC crossover region, where nanoscopic coexistence takes place. Surprisingly, though, this coexisting region seems to be immeasurably small or totally absent in La-1111.\cite{Luetkens2009,Khasanov2011}
Future measurements using Ce-1111 samples with finely tuned fluorine doping could be useful to better determine the extension of the M-SC coexisting region.

\begin{acknowledgments}
This work was performed at the Swiss Muon Source S$\mu$S, Paul Scherrer Institut 
(PSI), Switzerland and was partially  supported  by  MIUR  under  project PRIN2008 XWLWF9. The authors are grateful to A.~Amato for the instrumental support.
T.S.\ acknowledges support from the Schweizer Nationalfonds (SNF) and the NCCR program MaNEP.
\end{acknowledgments}

\appendix

\section{\label{ssec:SEM}SEM analysis}
\begin{figure}[th]
\centering
\vspace{5 mm}
\includegraphics[width=0.35\textwidth]{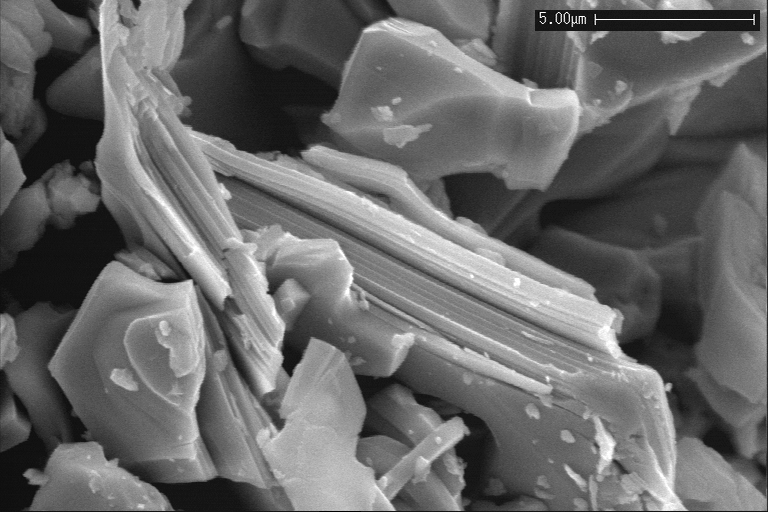}        
\caption{\label{fig:SEM}SEM image of $x=0.06$ sample showing the grain morphology.}
\end{figure}

To check for possible anomalies related to the varying fluorine content, detailed 
SEM analysis were performed on samples having different $x$ values. The morphology 
of the tested specimens turned out to be almost independent of doping. As an example, 
Fig.~\ref{fig:SEM} shows the SEM image of a sample with a nominal $x=0.06$.

\section{\label{ssec:LFmuSR}LF-$\mu$SR measurements}
The absence of oscillations in the $\mu$SR signal for fluorine dopings above $\sim 4\%$, could be due either to a wide distribution of static fields, or to strongly fluctuating (i.e.\ dynamic) magnetic moments. Although we could reasonably expect the static picture to reflect the physics of our system, we still carried out LF-$\mu$SR in the representative $x=0.06$ case. 
\begin{figure}[ht]
\centering
\includegraphics[width=0.4\textwidth]{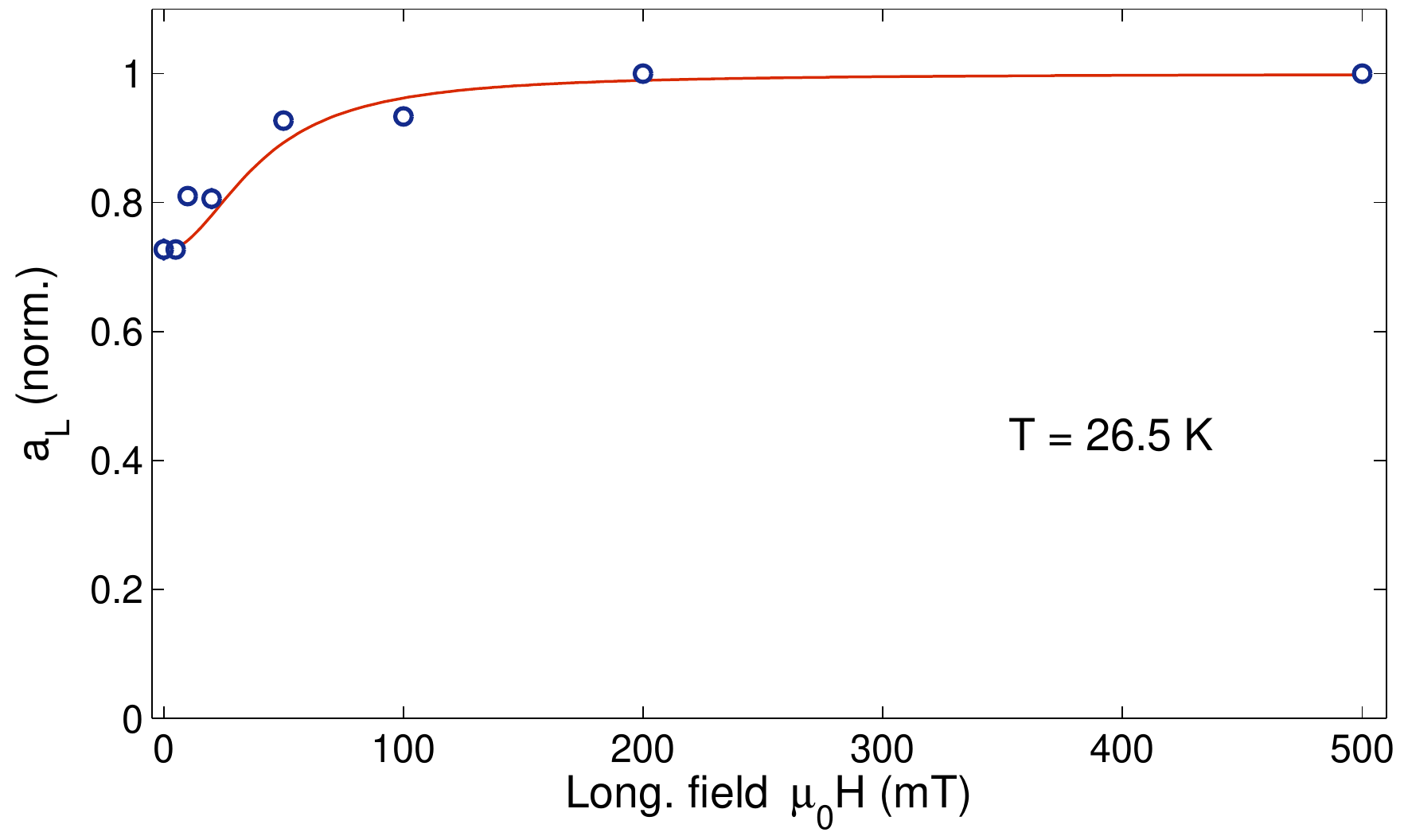}       
\caption{\label{fig:LFscan}(Color online) Longitudinal field scan at $T=26.5$~K for 
the $x=0.06$ sample. The solid line represents a fit using the polarization recovery function.\cite{Cox1987,Blundell2004}}
\end{figure}
In a so-called LF-decoupling experiment\cite{Hayano1979} an external magnetic field $B_{\parallel}$ is applied along the initial muon-spin direction. If $B_{\parallel}$ is of the same order of or higher than the internal static fields, then it will have a large influence on the muon polarization through a ``spin-locking'' effect. On the other hand, in case of strongly fluctuating internal fields the effect of the external field is barely noticeable.
By applying longitudinal fields in the range 0 to 500 mT, we could observe a clear polarization recovery for $B_{\parallel} \gtrsim 200$ mT (see Fig.~\ref{fig:LFscan}), in agreement with the hypothesis of a static distribution of the internal fields.


\end{document}